\documentclass[conference, letterpaper]{./IEEEtran}

\ifCLASSINFOpdf
  \usepackage[pdftex]{graphicx}
\else
 
\fi

\usepackage{amsmath}
\usepackage{amssymb}
\usepackage{amsfonts}

\usepackage{subfig}

\usepackage{multicol}

\begin{document}

\newtheorem{theorem}{Theorem}
\newtheorem{definition}{Definition}

\title{DVB-S2 Spectrum Efficiency Improvement with Hierarchical Modulation}

\author{\IEEEauthorblockN{Hugo M{\'e}ric\IEEEauthorrefmark{1}\IEEEauthorrefmark{2} and
Jos{\'e} Miguel Piquer\IEEEauthorrefmark{2}\\
}
\IEEEauthorblockA{\IEEEauthorrefmark{1}INRIA Chile, Santiago, Chile}
\IEEEauthorblockA{\IEEEauthorrefmark{2}NIC Chile Research Labs, Santiago, Chile\\
Email: hugo.meric@inria.cl, jpiquer@nic.cl}}

\maketitle

\begin{abstract}
We study the design of a DVB-S2 system in order to maximise spectrum efficiency. This task is usually challenging due to channel variability. Modern satellite communications systems such as DVB-SH and DVB-S2 rely mainly on a time sharing strategy to optimise the spectrum efficiency. Recently, we showed that combining time sharing with hierarchical modulation can provide significant gains (in terms of spectrum efficiency) compared to the best time sharing strategy. However, our previous design does not improve the DVB-S2 performance when all the receivers experience low or large signal-to-noise ratios. In this article, we introduce and study a hierarchical QPSK and a hierarchical 32-APSK to overcome the previous limitations. We show in a realistic case based on DVB-S2 that the hierarchical QPSK provides an improvement when the receivers experience poor channel condition, while the 32-APSK increases the spectrum efficiency when the receivers experience good channel condition.  
\end{abstract}

\IEEEpeerreviewmaketitle

\section{Introduction}

In most broadcast applications, the Signal-to-Noise Ratio (SNR) experienced by each receiver can be quite different. For instance, in satellite communications the channel quality decreases with the presence of clouds in Ku or Ka band, or with shadowing effects of the environment in lower bands. The first solution for broadcasting was to design the system for the worst-case reception, but this leads to poor performance as many receivers do not exploit their full potential. Two other schemes were proposed in \cite{cover} and \cite{bergmans}: time division multiplexing with variable coding and modulation, and superposition coding. Time division multiplexing, or time sharing, allocates a proportion of time to communicating with each receiver using any modulation and error protection level. This functionality, called Variable Coding and Modulation (VCM) \cite{s2}, is in practice the most used in standards today. If a return channel is available, VCM may be combined with Adaptive Coding and Modulation (ACM) to optimise the transmission parameters \cite{s2}. In superposition coding, the available energy is shared among several service flows which are sent simultaneously in the same band.
\begin{figure}[!t]
\centering
\includegraphics[width = 0.775\columnwidth]{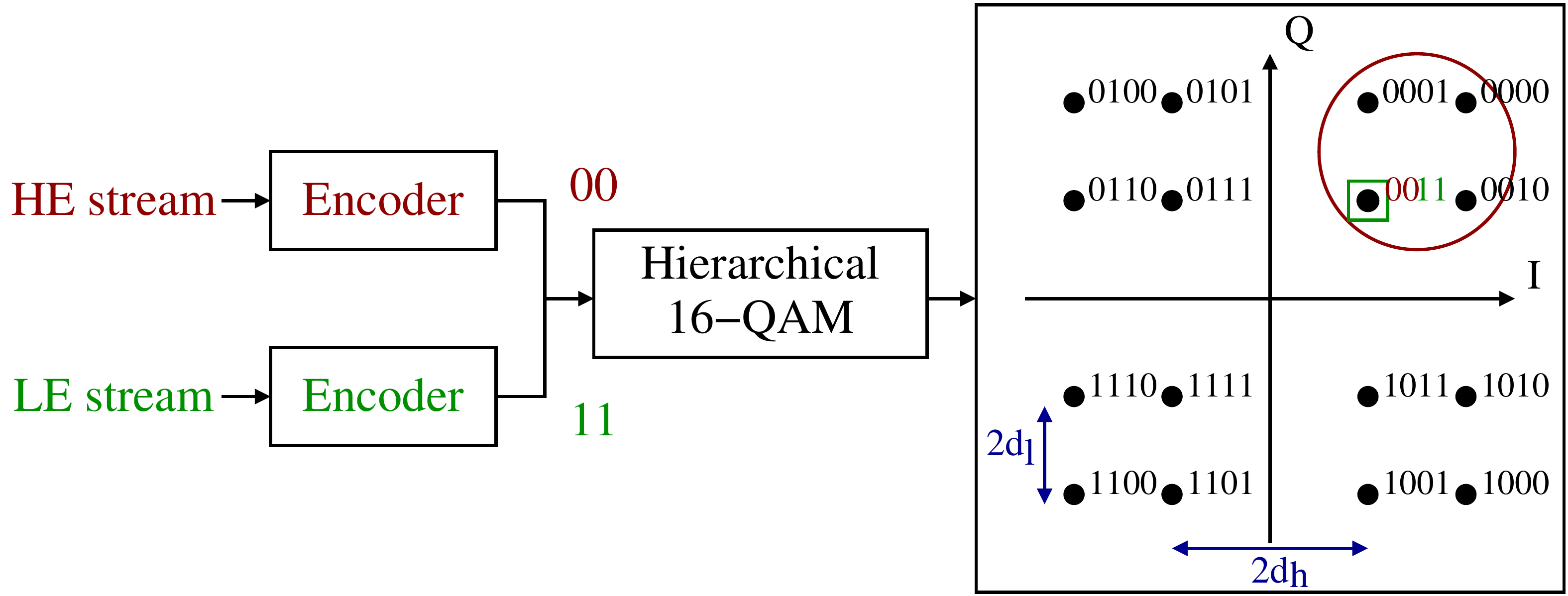}
\caption{Hierarchical modulation using a 16-QAM. Each constellation symbol carries data from 2 streams.}
\label{hm_principle}
\end{figure}
This scheme was introduced by Cover in \cite{cover} in order to improve the transmission rate from a single source to several receivers. When communicating with two receivers, the principle is to superimpose information for the user with the best SNR. This superposition can be done directly at the Forward Error Correction (FEC) level or at the modulation level as shown in \figurename~\ref{hm_principle} with a 16 Quadrature Amplitude Modulation (16-QAM).

Hierarchical modulation is a practical implementation of superposition coding. Although hierarchical modulation has been introduced to improve throughput, it has many others applications such as providing unequal protection \cite{svc_hm}, broadcasting local content \cite{local_content}, improving the performance of relay communication system \cite{relaycom} or backward compatibility \cite{backward_compatibility,broad05}. Note that none of the previous works use hierarchical modulation to improve the spectrum efficiency.

Our work focuses on using hierarchical modulation in modern broadcast systems to increase the transmission rate. For instance, even if the Low-Density Parity-Check (LDPC) codes of DVB-S2\footnote{Digital Video Broadcasting - Satellite - Second Generation} approach the Shannon limit for the Additive White Gaussian Noise (AWGN) channel with one receiver \cite{SAT:SAT787}, the throughput can be greatly increased for the broadcast case. Indeed, we recently showed that combining ACM with hierarchical modulation improves the spectrum efficiency of a DVB-S2 system \cite{broad13}. To that end, we used the hierarchical 16 Amplitude and Phase-Shift Keying (16-APSK) modulation. The performance improvement is significant, however there is no gain when all the receivers experience low or large SNRs. In this paper, we introduce and study a hierarchical Quadrature Phase-Shift Keying (QPSK) and a hierarchical 32-APSK to tackle these limitations. The main contribution is the presentation of  two novel hierarchical modulations to improve the spectrum efficiency of an AWGN broadcast channel.
 
The paper is organised as follows: Section~\ref{part2} introduces and studies two new hierarchical modulations. In Section~\ref{part3}, we show how hierarchical modulation can improve the performance of satellite communication systems. Section~\ref{part4} studies on a DVB-S2 use case the spectrum efficiency improvement with the previous modulations. Finally, Section~\ref{part5} concludes the paper by summarising the results and presenting the future work.

\section{Hierarchical modulation}\label{part2}

This part first introduces the principle of hierarchical modulation. Then we introduce the hierarchical QPSK and 32-APSK modulations. Finally, we evaluate with simulations the performance of both modulations on an AWGN channel.

\subsection{Hierarchical modulation}
As already mentioned, hierarchical modulations merge several streams in a same symbol. The available energy is shared between each stream. In this paper, two streams are considered. When hierarchical modulation is used for unequal protection purposes, these flows are called High Priority (HP) and Low Priority (LP) streams. However, unequal protection is not the goal of our work, so we will now refer to High Energy (HE) and Low Energy (LE) streams for the streams containing the most and the least energy, respectively.

As each stream usually does not use the same energy, hierarchical modulations often rely on \emph{non-uniform constellations} where the symbols are not uniformly distributed in the space. The geometry of non-uniform constellations is described using the constellation parameter(s). For instance, the hierarchical 16-QAM in \figurename~\ref{hm_principle} uses a non-uniform 16-QAM described by the constellation parameter $\alpha = d_h/d_l$, where $2d_h$ is the minimum distance between two constellation points carrying different HE bits and $2d_l$ is the minimum distance between any constellation point (see \figurename~\ref{hm_principle}). The energy ratio between the two streams is
\begin{equation}
\frac{E_{he}}{E_{le}} = (1+\alpha)^2 , 
\end{equation}
where $E_{he}$ and $E_{le}$ correspond to the amount of energy allocated to the HE and LE streams, respectively \cite{local_content}.

In the literature, the most common hierarchical modulations are the hierarchical 16-QAM and 8-PSK. We now introduce the hierarchical QPSK and 32-APSK modulations. For both modulations, we use an energy argument to choose the constellation parameter(s).

\subsection{Hierarchical QPSK and 32-APSK}

\subsubsection{Hierarchical QPSK}
\begin{figure}[!ht]
\centering
\includegraphics[width = 0.525\columnwidth]{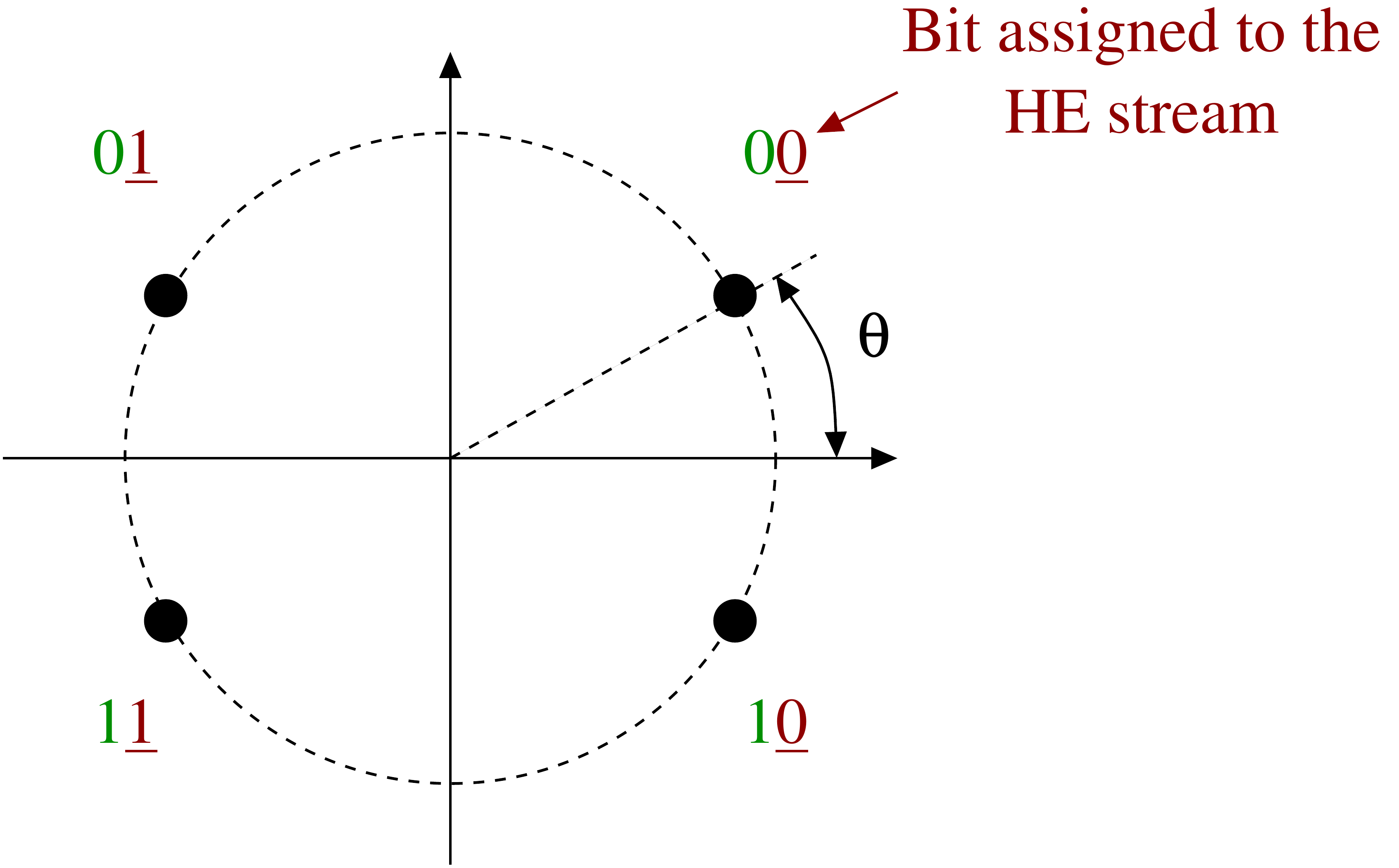}
\caption{Hierarchical QPSK. Constellation parameter: $\theta$}
\label{hqpsk}
\end{figure}

\figurename~\ref{hqpsk} shows the hierarchical QPSK. To describe the constellation geometry, it requires one parameter, $\theta$. Without loss of generality, we can assume that $0 \leqslant \theta \leqslant \pi/4$. The uniform QPSK corresponds to $\theta = \pi/4$. When $\theta < \pi/4$, the Least Significant Bit (LSB) in the mapping of each symbol is more protected than the Most Significant Bit (MSB) in the sense that its Bit Error Rate (BER) is smaller. The HE stream is transmitted with the LSB (see \figurename~\ref{hqpsk}).

For a given $\theta$ and energy per symbol ($E_s$), we have $E_{he}=\rho_{he}E_s$ where
\begin{equation}
\rho_{he} = \cos ( \theta )^2.
\label{energy_hqpsk}
\end{equation}
The parameter $\rho_{he}$ corresponds to the amount of energy allocated to the HE stream. Remark that $E_{le} = (1 - \rho_{he}) E_s$.

Even if the design of the hierarchical QPSK is easy, we show in Section~\ref{part4} that it improves the performance of a DVB-S2 system when the receivers experience poor channel conditions. To the best of our knowledge, the hierarchical QPSK has not been studied or used before.

\subsubsection{Hierarchical 32-APSK}
\begin{figure}[!ht]
\centering
\includegraphics[width = 0.7\columnwidth]{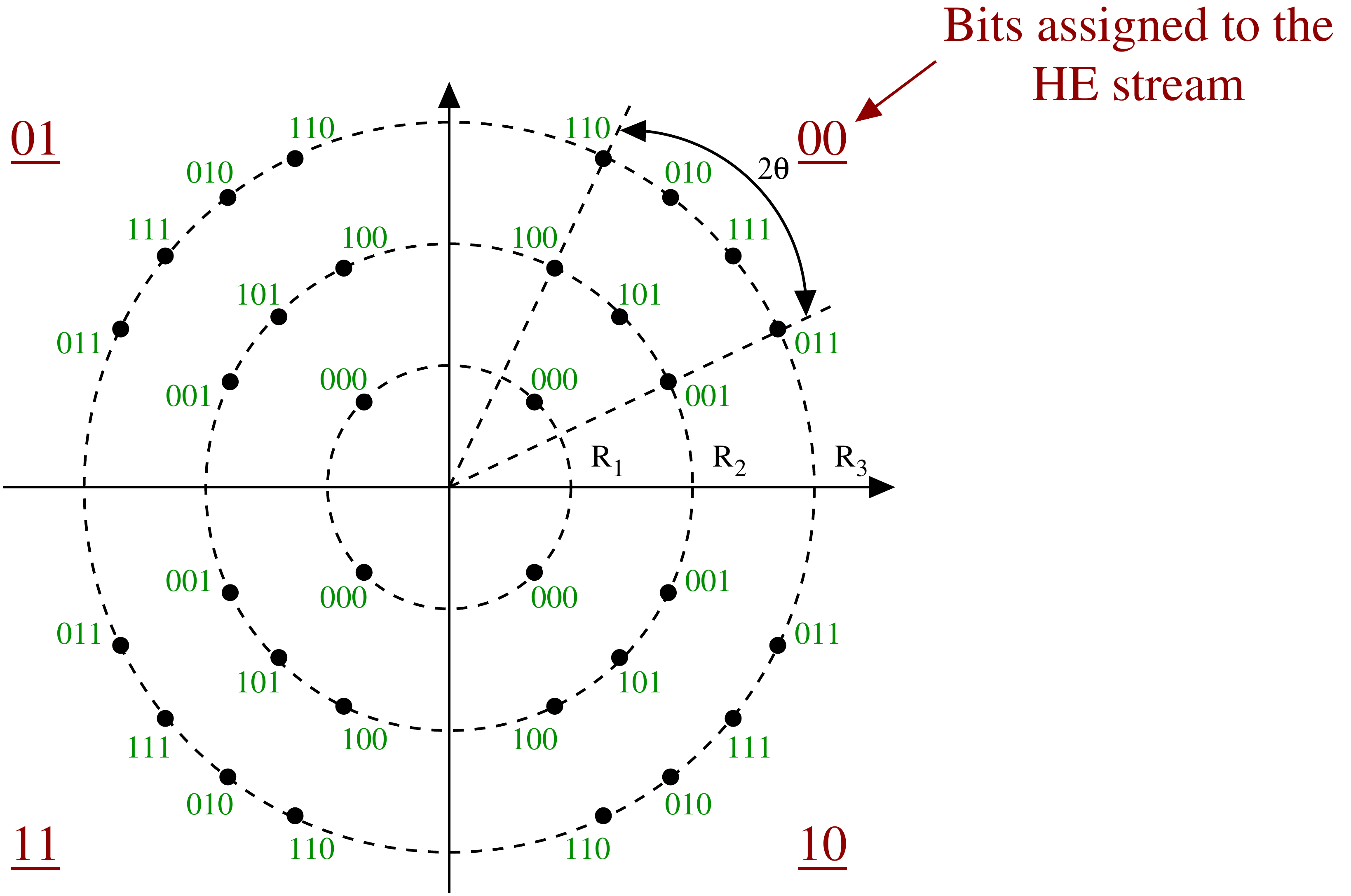}
\caption{Hierarchical 32-APSK. Constellation parameters: $\theta$, $\gamma_1=R_2/R_1$ and $\gamma_2=R_3/R_1$}
\label{32apsk}
\end{figure}

\figurename~\ref{32apsk} presents the hierarchical 32-APSK and the mapping used in this paper. The mapping is based on the Gray mapping of a 32-QAM. In each quadrant, the two MSB are identical (underlined bits in \figurename~\ref{32apsk}). These bits serve to transmit the HE stream. The constellation parameters are the ratio between the radius of the middle ($R_2$) and inner ($R_1$) rings $\gamma_1 = R_2/R_1$ , the ratio between the radius of the outer ($R_3$) and inner ($R_1$) rings $\gamma_2 = R_3/R_1$ and the half angle between the points on the outer ring in each quadrant $\theta$ (see \figurename~\ref{32apsk}).

We consider the energy of the HE stream, given by the energy of a QPSK modulation where the constellation points are located at the barycenter of the eight points in each quadrant. Using the polar coordinates, the barycenter in the upper right quadrant is
\begin{equation}
\textstyle
e^{i\pi/4} \frac{R_1+R_2 (1+e^{i\theta}+e^{-i\theta}) + R_3(e^{i\theta}+e^{-i\theta}+e^{i\theta/3}+e^{-i\theta/3})}{8} .
\label{barycentre}
\end{equation}

Moreover, the symbol energy $E_s$ is expressed as
\begin{equation}
E_s = \frac{4R_1^2+12R_2^2+16R_3^2}{32} = \frac{1+3\gamma_1^2+4\gamma_2^2}{8}R_1^2.
\label{Es}
\end{equation}

Combining (\ref{barycentre}) and (\ref{Es}), the distance $d$ of the barycenter to the origin is
\begin{equation}
d = \frac{1+ \gamma_1 \left(1+2 \cos\theta\right) + 2\gamma_2\left(\cos\theta +\cos\theta/3\right) }{\sqrt{8(1+3\gamma_1^2+4\gamma_2^2)}}\sqrt{E_s}.
\label{module}
\end{equation}

Finally, the energy of the HE stream is equals to $d^2$ which is the energy of a QPSK modulation with each constellation point located at a distance $d$ from the origin. Thus we can write $E_{he}=\rho_{he}E_s$ where
\begin{equation}
\rho_{he} = \frac{ \left( 1+ \gamma_1 \left(1+2 \cos\theta\right) + 2\gamma_2\left(\cos\theta +\cos\theta/3\right) \right)^2 }{8(1+3\gamma_1^2+4\gamma_2^2)}.
\label{energie}
\end{equation}

\subsubsection{Remark}
Equations (\ref{energy_hqpsk}) and (\ref{energie}) introduce $\rho_{he}$ as the ratio between the energy of the HE stream $E_{he}$ and the symbol energy $E_s$. As the HE stream contains more energy than the LE stream, we verify that $\rho_{he} \geqslant 0.5$.

\subsection{Performance of the hierarchical QPSK and 32-APSK}
In practical systems, several values of $\rho_{he}$ have to be chosen. For the hierarchical QPSK, a given $\rho_{he}$ corresponds to one value of $\theta$ as shown in (\ref{energy_hqpsk}). Table~\ref{adopted_pairs_hqpsk} resumes the adopted constellation parameters ($\theta$ is expressed in degree).
\begin{table}[ht!]
\renewcommand{\arraystretch}{1.1}
\caption{Constellation parameters for the hierarchical QPSK}
\label{adopted_pairs_hqpsk}
\centering
\begin{tabular}{c||c|c|c|c|c|c|c|c|c} 
\hline
$\rho_{he}$ & 0.5 & 0.55 & 0.6 & 0.65 & 0.7 & 0.75 & 0.8 & 0.85 & 0.9 \\
\hline
$\theta$ & 45 & 42 & 39 & 36 & 33 & 30 & 27 & 24 & 18 \\
\hline 
\end{tabular}
\end{table}

For the hierarchical 32-APSK, once the value of $\rho_{he}$ is known, there remains to pick one $(\gamma_1, \gamma_2, \theta)$ triple. We solve (\ref{energie}) with Matlab and choose the triple that minimises the decoding threshold of the HE stream averaged over all the DVB-S2 code rates. To obtain a fast evaluation of the decoding thresholds in function of the constellation parameters, we use the method described in \cite{wts}. Table~\ref{adopted_pairs} presents the adopted triples.
\begin{table}[ht!]
\renewcommand{\arraystretch}{1.1}
\caption{Hierarchical 32-APSK constellation parameters}
\label{adopted_pairs}
\centering
\begin{tabular}{c||c|c|c|c|c} 
\hline
$\rho_{he}$ & 0.7  & 0.75 & 0.8  & 0.85 & 0.9 \\
\hline
$\gamma_1$  & 2.4  & 1.8  & 1.6  & 1.6  & 1.8 \\
\hline
$\gamma_2$  & 5    & 3.4  & 2.6  & 2.2  & 2.4 \\
\hline
$\theta$    & 32.3 & 30.2 & 28.4 & 25.6 & 17.4 \\
\hline 
\end{tabular}
\end{table}

\begin{figure}[!ht]
\centering
\includegraphics[width = 0.875\columnwidth]{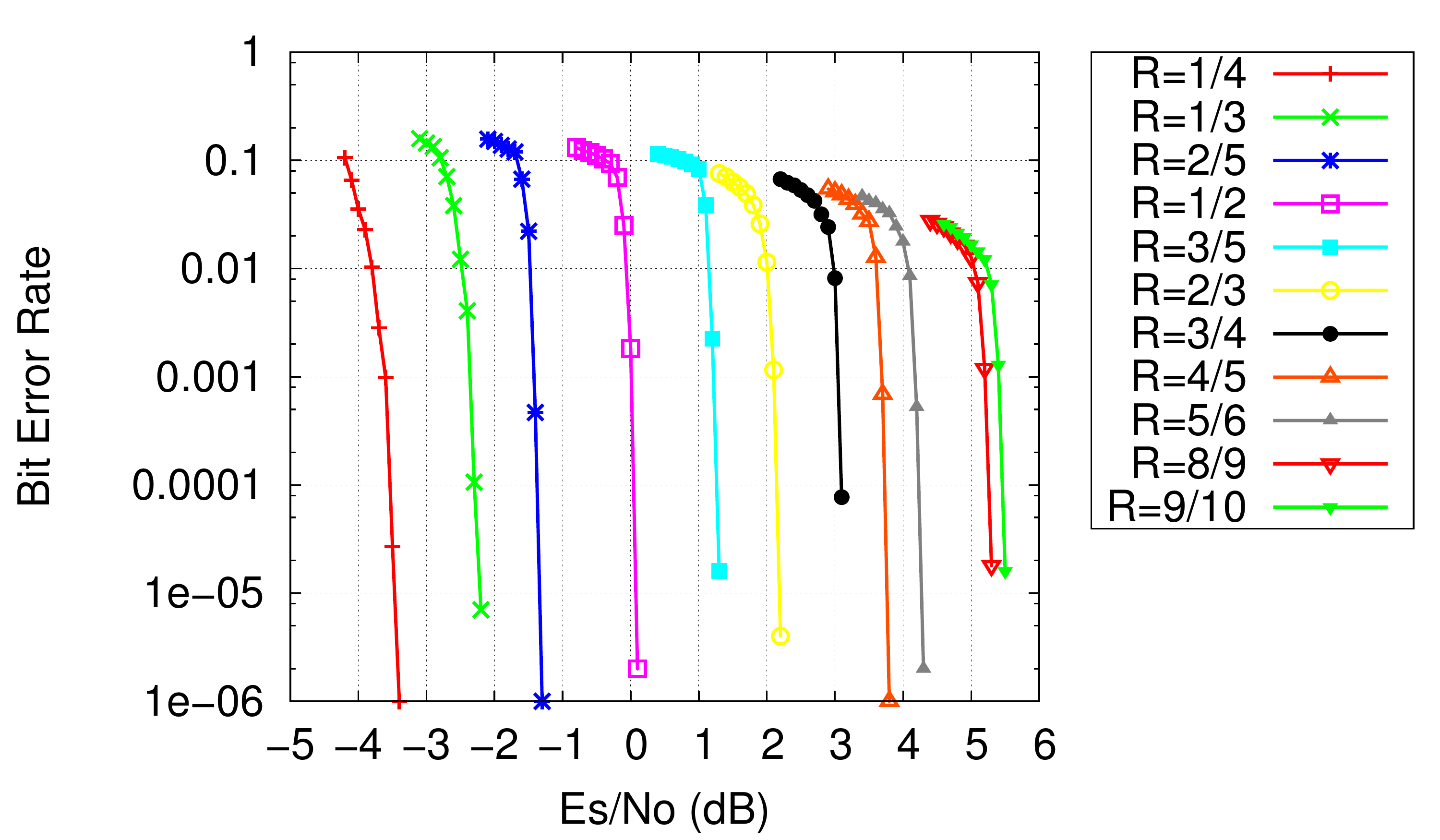}
\caption{BER for the hierarchical QPSK HE stream ($\rho_{he}=0.6$)}
\label{perf_ber}
\end{figure}

Finally, the performance in terms of BER of the hierarchical QPSK and 32-APSK are evaluated with simulations. We use the Coded Modulation Library \cite{cml} that already implements the DVB-S2 LDPC. The LDPC codewords are 64 800 bits long (normal FEC frame) and the iterative decoding stops after 50 iterations if no valid codeword has been decoded. Moreover, in our simulations, we wait until 10 decoding failures before computing the BER. If the BER is less than $10^{-4}$, then we stop the simulation. Our stopping criterion is less restrictive than in \cite{SAT:SAT787} (i.e., a packet error rate of $10^{-7}$) because simulations are time consuming. However, our simulations are sufficient to detect the waterfall region of the LDPC and then the performance of the code. For instance, \figurename~\ref{perf_ber} presents the BER of the HE stream for the hierarchical QPSK with $\rho_{he}=0.6$.

\section{Hierarchical modulation for Satcom systems}\label{part3}

In \cite{broad13}, we compared the two following schemes: time sharing with or without hierarchical modulation, referred to as hierarchical modulation and classical time sharing, respectively. This part reviews how the combination of ACM and hierarchical modulation (i.e., hierarchical modulation time sharing) improves the performance of a DVB-S2 system based on classical time sharing. A more detailed description can be found in \cite{broad13}. In this section and the rest of the paper, we consider an AWGN channel. We assume that the transmitter has knowledge of the SNR at the receivers. A concrete example is a DVB-S2 system that implements ACM.

\subsection{Case with two receivers}
We first consider one source communicating with two receivers, each one with a particular signal-to-noise ratio $SNR_i$ ($i=1,2$). Given $SNR_i$ and the transmission parameters, receiver $i$ has a spectrum efficiency $R_i$ which is the amount of useful data transmitted on the link over a given bandwidth. For instance, if the receiver successfully decodes a signal modulated with a QPSK and a coding rate of 1/3, then its spectrum efficiency equals $2 \times 1/3$ bit/s/Hz. In our study, the physical layer is based on the DVB-S2 standard \cite{dvbs2}.

When communicating with two receivers, non-hierarchical modulations achieve spectrum efficiencies of the form $(R_1,0)$ or $(0,R_2)$. Hierarchical modulation allows spectrum efficiencies of the form $(R_1,R_2)$ as the transmitted symbols carry information to both receivers. When two spectrum efficiencies pairs $(R_1,R_2)$ and $\left( R_1^*,R_2^* \right)$ are available, the time sharing strategy achieves any spectrum efficiency pair 
\begin{equation}
\left(\tau R_1+(1-\tau)R_1^* , \tau R_2+(1-\tau)R_2^* \right),
\end{equation}
where $0 \leqslant \tau \leqslant 1$ is the fraction of time allocated to $(R_1,R_2)$. Thus, if a set $\chi = \left\{ (R_1^i,R_2^i) | i=1,..,k \right\}$ of spectrum efficiencies pairs is available, any pair $(R_1, R_2)$ inside the convex hull of $\chi$ can be achieved.

In this paper, we are interested in \emph{offering the same (time-averaged) spectrum efficiency to all the receivers}. We note $R_{hm}$ and $R_{ts}$ as the spectrum efficiencies offered to both receivers by the hierarchical modulation and classical time sharing strategy, respectively. In \cite{broad13}, the receiver with the best SNR decodes the LE stream, while the receiver with the worst SNR decodes the HE stream. However, when the SNRs of both receivers are close, it is also interesting that the receiver with the best SNR decodes the HE stream. We consider both cases thereafter.

\begin{figure}[!ht]
\centering
\includegraphics[width = 0.975\columnwidth]{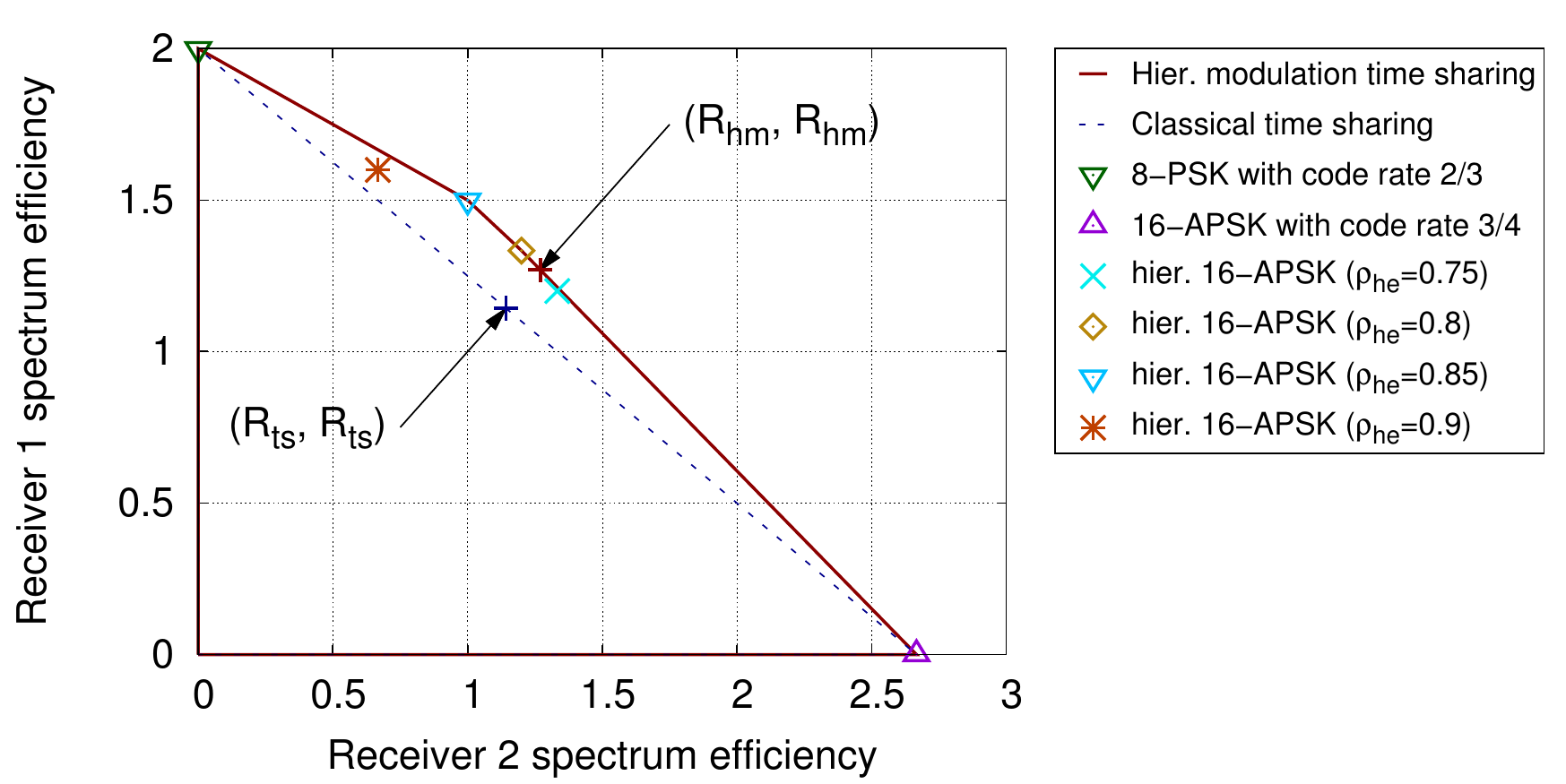}
\caption{Spectrum efficiency for 2 receivers ($SNR_1=7$ dB, $SNR_2=$10 dB)}
\label{rate_region}
\end{figure}

\figurename~\ref{rate_region} shows on an example the spectrum efficiency improvement when the receivers experience a SNR of 7 and 10 dB (to lighten \figurename~\ref{rate_region}, only the hierarchical 16-APSK is used). 

\subsection{Case with $n$ receivers}
We consider a broadcast system with $n$ receivers. When transmissions do not involve hierarchical modulation , we assume that receiver $i$ has a spectrum efficiency $R_i$ which corresponds to the best spectrum efficiency it can manage. Then the classical time sharing offers the following spectrum efficiency to all the receivers
\begin{equation}
R_{ts} = \left( \sum_{j=1}^n \frac{1}{R_j}\right)^{-1}.
\end{equation}

For the hierarchical modulation time sharing, the first step is to group the receivers in pairs in order to use hierarchical modulation. Many possibilities are available. We adopt the following strategy: from any set of receivers, we pick the two receivers with the largest SNR difference, group them and repeat this operation. We showed in \cite{broad13} that this strategy generally leads to significant performance improvement. 

Once the pairs have been chosen, we compute for each pair the spectrum efficiency as previously described with two receivers. We note $R_{hm,i}$ the spectrum efficiency for each pair ($1 \leqslant i \leqslant n/2$). As the terms $R_{hm,i}$ are different, we need to equalise the spectrum efficiency between each receiver. This is done using time sharing. Finally, the spectrum efficiency is
\begin{equation}
R_{hm} = \left( \sum_{j=1}^{n/2} \frac{1}{R_{hm, j}}\right)^{-1}.
\end{equation}

In \cite{broad13}, we compare the values of $R_{hm}$ and $R_{ts}$ for various scenarios. Using the grouping strategy presented above, we achieved significant gains (up to 10\%) with the hierarchical modulation time sharing. However, there is no gain when all the receivers experience low or large SNRs.

\section{Performance evaluation}\label{part4}
To evaluate the performance of the two hierarchical modulations proposed in Section~\ref{part2} for real systems, we first present a model to estimate the SNR distribution of the receivers for an AWGN channel. Then we evaluate the performance of hierarchical modulation time sharing and discuss the spectrum efficiency improvement of each modulation.

\subsection{Channel model}
\begin{figure}[!ht]
\centering
\includegraphics[width = 0.375\columnwidth]{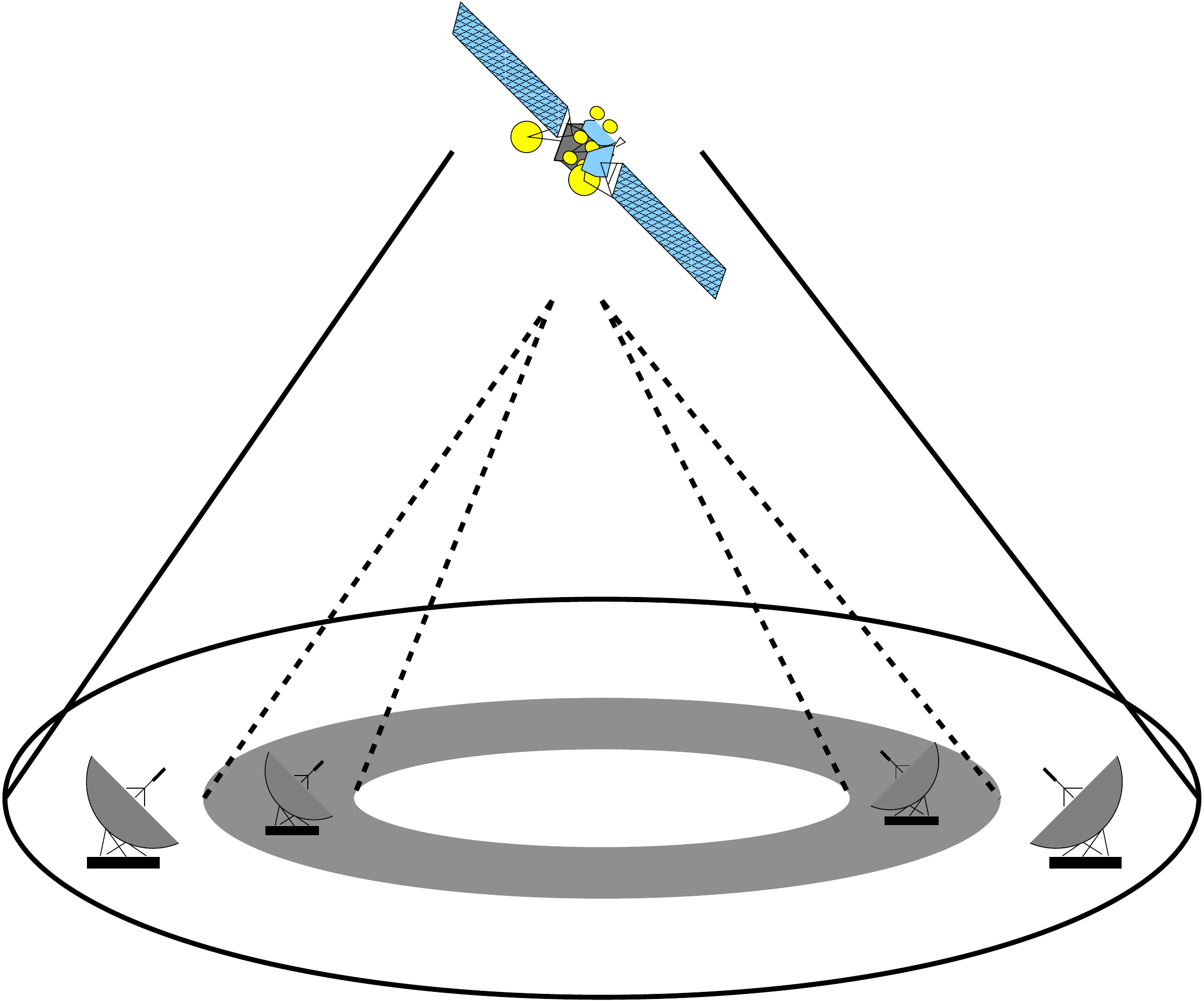}
\caption{Satellite broadcasting area}
\label{spot_sat}
\end{figure}

We consider the set of receivers located in a given spot beam of a geostationary satellite broadcasting in the Ka band. The model takes into account two main sources of attenuation: the relative location of the terminal with respect to the center of (beam) coverage and the weather. Concerning the attenuation due to the location, the idea is to set the SNR at the center of the spot beam $\text{SNR}_{max}$ and use the radiation pattern of a parabolic antenna to model the attenuation. An approximation of the radiation pattern is 
\begin{equation}
G(\theta) = G_{max} \left( 2\frac{J_1 \left( \sin(\theta) \frac{\pi D}{\lambda} \right)}{\sin(\theta) \frac{\pi D}{\lambda}} \right)^2,
\label{eq_rayonnement}
\end{equation}
where $J_1$ is the first order Bessel function, $D$ is the antenna diameter and $\lambda=c/f$ is the wavelength \cite{antenna}. Our simulations use $D=1.5\text{ m}$ and $f=20 \text{ GHz}$. Moreover, we consider a typical multispot system where the edge of each spot beam is 4 dB below the center of coverage. Assuming a uniform repartition of the population, the proportion of the receivers experiencing an attenuation between two given values is the ratio of the ring area over the disk as shown in \figurename~\ref{spot_sat}. The ring area is computed knowing the satellite is geostationary and using (\ref{eq_rayonnement}).

\begin{figure}[!ht]
\centering
\includegraphics[width = 0.9\columnwidth]{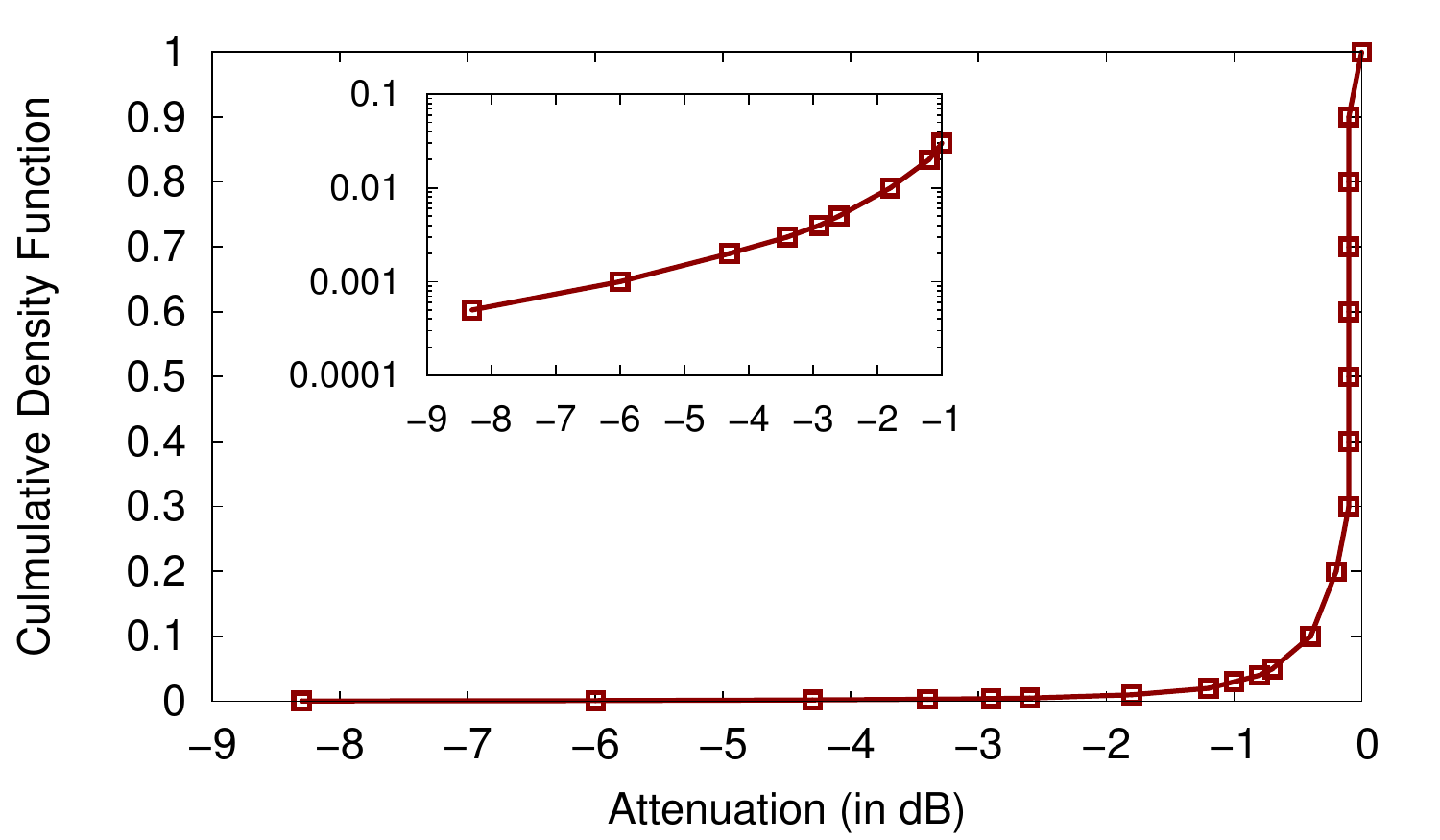}
\caption{Attenuation distribution (due to weather)}
\label{distrib_s2}
\end{figure}

\figurename~\ref{distrib_s2}, provided by the Centre National d'Etudes Spatiales (CNES), presents the attenuation distribution of the Broadcasting Satellite Service (BSS) band. More precisely, it is a temporal distribution for a given location in Toulouse, France. In this paper, we assume that the SNR distribution for the receivers in the beam coverage at a given time is equivalent to the temporal distribution at a given location.

Finally, our model combines the two attenuations previously described to estimate the SNR distribution. From a set of receivers, we first compute the attenuation due to the location. Then, for each receiver we draw the attenuation caused by the weather according to the distribution in \figurename~\ref{distrib_s2}.

\begin{figure*}[!ht]
\centering
\includegraphics[width = 0.75\textwidth]{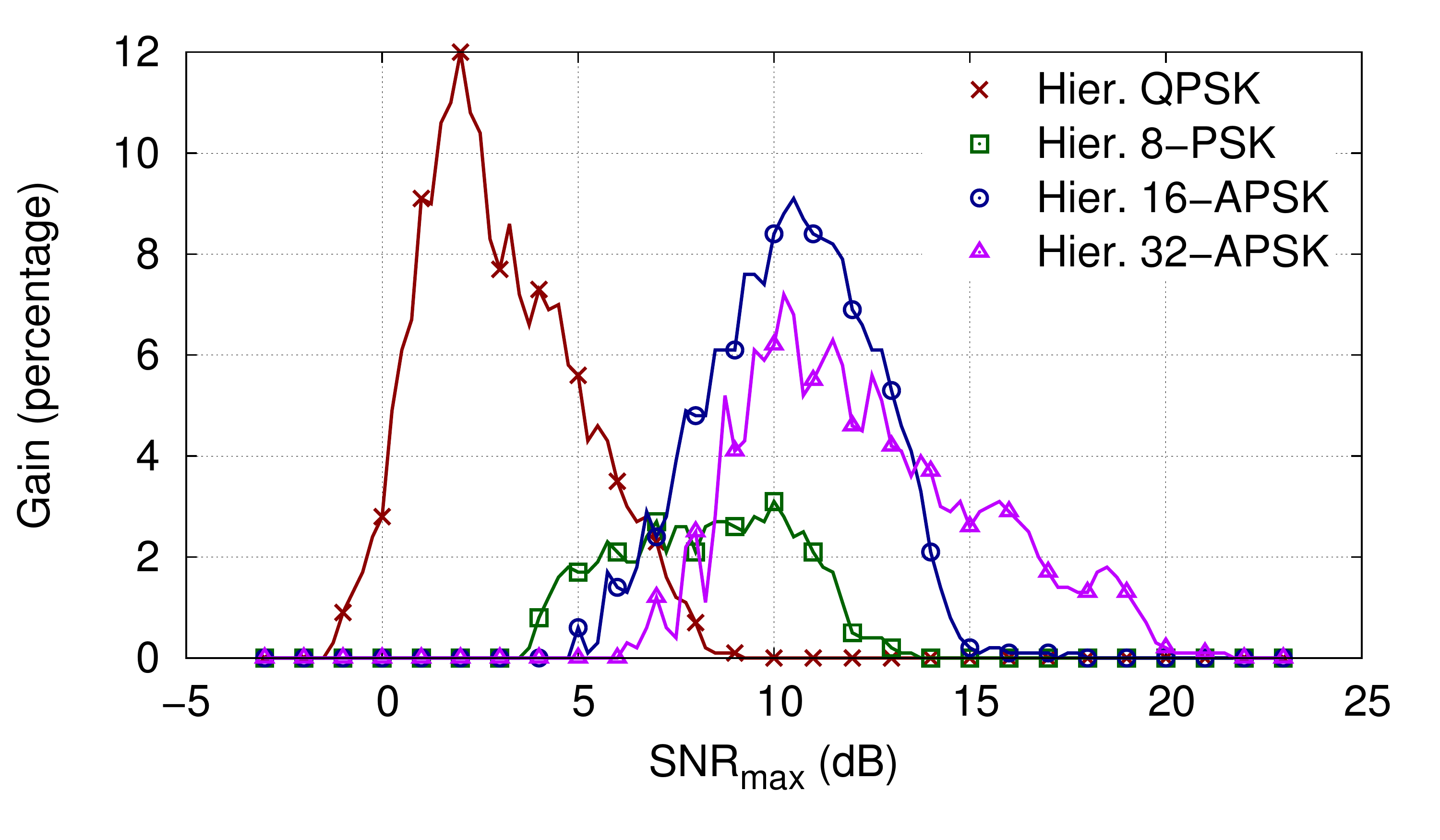}
\caption{Spectrum efficiency gains of the hierarchical modulation time sharing with 500 receivers. Each curve is an average over 100 simulations. The gains of the hierarchical 16-APSK are consistent with \cite[Fig. 9]{broad13}.}
\label{results}
\end{figure*}

\subsection{Transmission parameters for the simulations}
Our simulations involve the following modulations: QPSK, 8-PSK, 16-APSK, 32-APSK and their hierarchical versions. The error performance of the non-hierarchical modulations are summarised in \cite[Table 13]{dvbs2}. The hierarchical QPSK and 32-APSK have been introduced in Section~\ref{part2} and their performance are given in the Appendix. We use the four hierarchical 16-APSK modulations studied in \cite{broad13}. For the hierarchical 8-PSK, the constellation parameter $\theta$ is defined as the half angle between two points in the same quadrant. We chose $\theta = 30^\circ,27^\circ,24^\circ,18^\circ$ for the simulations.

\subsection{Simulations results}
\figurename~\ref{results} presents the gains (in terms of spectrum efficiency) of hierarchical modulation time sharing compared to classical time sharing for a broadcasting area with 500 receivers. For each simulation, the SNR value of each receiver is drawn according to the distribution presented above. Note that this SNR is fixed over all times for a given simulation. We also assume that the transmitter has knowledge of the SNR at the receivers. In practice, this corresponds to a system that implements ACM. For one system configuration (i.e., the parameter $SNR_{max}$ is set), we present the average gains over 100 simulations. Moreover, we show the results independently for each hierarchical modulation in order to visualise how it affects performance.

For low SNR configuration, the hierarchical QPSK is the only modulation that improves the performance. For instance, there is a gain of 9\% for $SNR_{max}=1$ dB and we observe an improvement up to 12\% for $SNR_{max} \approx 2$ dB. However, when $SNR_{max} \geqslant 8$ dB, the hierarchical QPSK does not increase the spectrum efficiency anymore and we need to increase the modulation order. 

For large SNR configuration, both the hierarchical 16-APSK and 32-APSK improve the performance. When $5 \text{ dB } \leqslant SNR_{max} \leqslant 13$ dB, the hierarchical 16-APSK obtains a slight advantage. However, the results point out that the hierarchical 32-APSK performs better than the hierarchical 16-APSK for $SNR_{max} \geqslant 13 $ dB. Indeed, the hierarchical 32-APSK still offers some gains, e.g., around 3\% for $SNR_{max}=16$ dB.  

Unlike the others modulations, the hierarchical 8-PSK does not provide any significant gain. Moreover, either the hierarchical QPSK ($SNR_{max} \leqslant 6.5 $ dB) or the hierarchical 16-APSK ($SNR_{max} \geqslant 6.5 $ dB) outperforms the hierarchical 8-PSK. When all the modulations are used (the curve is not presented to make \figurename~\ref{results} easier to read), the gains follow the curve of the QPSK for $SNR_{max} \leqslant 6 $ dB, then the curve of the 16-APSK until $SNR_{max} = 13$ dB and finally the curve of the 32-APSK. We just observe a slight improvement when $SNR_{max}$ is around 6 dB.

In practical systems, the transmission parameters (modulation and coding rates) have to be signaled. In this paper, the simulations involve 22 hierarchical modulations and there are 11 code rates. Thus it requires 12 bits of signalisation ($\log_2 (11 \times 11 \times 22) \approx 11.4$). However, it is possible to reduce that number. For instance, the hierarchical 8-PSK is not useful and can be removed. To avoid completely the signaling of the modulation, a solution is to use modulation recognition that consists in identifying at the receiver the modulation used by the transmitter \cite{reconnaissance_modulation}. However, it is necessary to verify that this does not decrease too much the performance and that the complexity at the receiver is still acceptable. Another solution to reduce the number of transmission parameters is investigated in \cite{subset_optimization}.

\begin{table*}[!ht]
\renewcommand{\arraystretch}{1.05}
\setlength{\tabcolsep}{0.13cm}
\caption{Hierarchical QPSK decoding thresholds (in dB)}
\label{decoding_threshold}
\centering
\begin{tabular}{c||c|c|c|c|c|c|c|c|c|c|c|c|c|c|c|c|c} 
\hline
Code & $\rho_{he}=0.5$ & \multicolumn{2}{c|}{$\rho_{he}=0.55$} & \multicolumn{2}{c|}{$\rho_{he}=0.6$} & \multicolumn{2}{c|}{$\rho_{he}=0.65$} & \multicolumn{2}{c|}{$\rho_{he}=0.7$} & \multicolumn{2}{c|}{$\rho_{he}=0.75$} & \multicolumn{2}{c|}{$\rho_{he}=0.8$} & \multicolumn{2}{c|}{$\rho_{he}=0.85$} & \multicolumn{2}{c}{$\rho_{he}=0.9$} \\ 
\cline{2-18}
rate & HE/LE & HE & LE & HE & LE & HE & LE & HE & LE & HE & LE & HE & LE & HE & LE & HE & LE \\
\hline
1/4 & -2.6 & -3.1 & -2.2 & -3.5 & -1.7 & -3.8 & -1 & -4.2 & -0.4 & -4.4 & 0.4 & -4.7 & 1.2 & -4.9 & 2.1 & -5.2 & 4.5 \\
\hline
1/3 & -1.4 & -1.8 & -0.9 & -2.2 & -0.4 & -2.6 & 0.2 & -2.9 & 0.9 & -3.2 & 1.6 & -3.4 & 2.5 & -3.6 & 3.4 & -4 & 5.8 \\
\hline 
2/5 & -0.5 & -1 & 0 & -1.3 & 0.5 & -1.7 & 1.1 & -2 & 1.8 & -2.3 & 2.5 & -2.5 & 3.3 & -2.8 & 4.3 & -3.1 & 6.7 \\
\hline 
1/2 & 0.9 & 0.5 & 1.4 & 0.1 & 1.2 & -0.3 & 2.5 & -0.6 & 3.2 & -0.9 & 3.9 & -1.1 & 4.8 & -1.3 & 5.7 & -1.7 & 8.1 \\
\hline 
3/5 & 2.1 & 1.7 & 2.6 & 1.3 & 3.1 & 1 & 3.7 & 0.7 & 4.4 & 0.4 & 5.1 & 0.1 & 6 & -0.1 & 6.9 & -0.4 & 9.3 \\
\hline 
2/3 & 3 & 2.6 & 3.5 & 2.2 & 4 & 1.8 & 4.6 & 1.5 & 5.3 & 1.3 & 6 & 1 & 6.9 & 0.8 & 7.8 & 0.4 & 10.2 \\
\hline 
3/4 & 4 & 3.5 & 4.5 & 3.1 & 5 & 2.8 & 5.6 & 2.5 & 6.3 & 2.2 & 7 & 2 & 7.8 & 1.8 & 8.8 & 1.4 & 11.2 \\
\hline 
4/5 & 4.6 & 4.2 & 5.1 & 3.8 & 5.6 & 3.4 & 6.2 & 3.1 & 6.9 & 2.8 & 7.6 & 2.6 & 8.5 & 2.4 & 9.4 & 2 & 11.8 \\
\hline 
5/6 & 5.1 & 4.7 & 5.6 & 4.3 & 6.1 & 3.9 & 6.7 & 3.6 & 7.4 & 3.3 & 8.1 & 3.1 & 9 & 2.9 & 9.9 & 2.5 & 12.3 \\
\hline 
8/9 & 6.1 & 5.7 & 6.6 & 5.3 & 7.1 & 5 & 7.7 & 4.7 & 8.4 & 4.4 & 9.1 & 4.1 & 10 & 3.9 & 10.9 & 3.6 & 13.3 \\
\hline 
9/10 & 6.3 & 5.9 & 6.8 & 5.5 & 7.4 & 5.2 & 7.9 & 4.9 & 8.6 & 4.6 & 9.4 & 4.3 & 10.2 & 4.1 & 11.1 & 3.8 & 13.5 \\
\hline   
\end{tabular}
\end{table*}

Recently, a new method called Bit Division Multiplexing (BDM) has also been proposed to increase the throughput of broadcasting systems \cite{6509443}. In fact, hierarchical modulation is a special case of BDM. In \cite{6509443}, the authors study BDM from a theoretical point of view and use uniform constellations. We strongly believe that BDM combined with non-uniform constellations could further improve the performance of satellite communication systems.

\section{Conclusion}\label{part5}

We introduce and study the hierarchical QPSK and the hierarchical 32-APSK. We propose these two modulations to increase the spectrum efficiency offered to the receivers in a DVB-S2 system. To the best of our knowledge, these modulations have not been studied before. Here, we chose the constellation parameters according to an energy argument. Then we showed that hierarchical QPSK achieves significant gains for low SNRs (up to 12\%). The hierarchical 32-APSK is efficient for large SNRs but the gains are less important.

Future work will explore a better computation of the 32-APSK constellation parameters to obtain higher gains for large SNRs. We also plan to investigate the combination of non-uniform constellations with bit division multiplexing to further increase the spectrum efficiency of satellite broadcasting systems.

\appendix\label{appendix}

Table~\ref{decoding_threshold} and \ref{decoding_threshold2} summarise the decoding thresholds (obtained by simulations) to achieve a BER of $10^{-4}$ for the hierarchical QPSK and 32-APSK over an AWGN channel. 

\begin{table}[!ht]
\renewcommand{\arraystretch}{1.05}
\setlength{\tabcolsep}{0.13cm}
\caption{Hierarchical 32-APSK decoding thresholds (in dB)}
\label{decoding_threshold2}
\centering
\begin{tabular}{c||c|c|c|c|c|c|c|c|c|c} 
\hline
Code & \multicolumn{2}{c|}{$\rho_{he}=0.7$} & \multicolumn{2}{c|}{$\rho_{he}=0.75$} & \multicolumn{2}{c|}{$\rho_{he}=0.8$} & \multicolumn{2}{c|}{$\rho_{he}=0.85$} &  \multicolumn{2}{c}{$\rho_{he}=0.9$} \\ 
\cline{2-11}
rate & HE & LE  & HE & LE & HE & LE & HE & LE & HE & LE \\
\hline
1/4 & 0 & 6 & -0.6 & 6.6 & -1.1 & 7.6 & -1.6 & 9 & -1.9 & 10.5 \\
\hline
1/3 & 1.7 & 7.3 & 0.9 & 8 & 0.4 & 9 & -0.1 & 10.6 & -0.6 & 12 \\
\hline 
2/5 & 3 & 8.5 & 2.2 & 9.2 & 1.5 & 10.2 & 0.9 & 11.7 & 0.4 & 13 \\
\hline 
1/2 & 5.2 & 10.1 & 4.2 & 10.8 & 3.4 & 11.8 & 2.7 & 13.3 & 2 & 14.7 \\
\hline 
3/5 & 7.5 & 11.5 & 6.2 & 12.3 & 5.1 & 13.2 & 4.3 & 14.6 & 3.4 & 16.2 \\
\hline 
2/3 & 9 & 12.5 & 7.5 & 13.3 & 6.4 & 14.2 & 5.4 & 15.6 & 4.5 & 17.2 \\
\hline 
3/4 & 11 & 13.6 & 9.3 & 14.4 & 8 & 15.2 & 6.8 & 16.6 & 5.8 & 18.2 \\
\hline 
4/5 & 12.3 & 14.4 & 10.4 & 15.2 & 9 & 15.9 & 7.8 & 17.2 & 6.6 & 18.9 \\
\hline 
5/6 & 13.3 & 15.1 & 11.3 & 15.9 & 9.9 & 16.5 & 8.5 & 17.7 & 7.2 & 19.5 \\
\hline 
8/9 & 15.4 & 16.4 & 13.2 & 17.2 & 11.6 & 17.6 & 10.2 & 18.7 & 8.7 & 20.6 \\
\hline 
9/10 & 15.9 & 16.6 & 13.6 & 17.4 & 12 & 17.9 & 10.5 & 18.9 & 9 & 20.8 \\
\hline   
\end{tabular}
\end{table}

\ifCLASSOPTIONcaptionsoff
  \newpage
\fi

\nocite{*}
\bibliographystyle{IEEEtran}
\bibliography{biblio}

\end{document}